\documentclass[showpacs,preprintnumbers,amsmath,amssymb]{revtex4}

\usepackage{graphicx}
\usepackage{dcolumn}
\usepackage{bm}

\begin{document}

\preprint{}

\title{Quantum logic networks for probabilistic teleportation of many particle state of general form}

\author{Ting Gao, $^{1,2}$ Fengli Yan $^{3,4}$ and Zhixi Wang $^1$}
 \affiliation{%
  $^1$ Department of Mathematics, Capital Normal University, Beijing 100037,
 China\\
 $^2$ College of Mathematics and Information Science, Hebei Normal
University, Shijiazhuang
050016, China\\
$^3$ Department of Physics, Hebei Normal University, Shijiazhuang 050016, China\\
$^4$ CCAST (World Laboratory), P.O. Box 8730, Beijing 100080,
China}%

\begin{abstract}
The scheme for probabilistic teleportation of an N-particle state of general form is proposed. As the special
cases we construct efficient quantum logic networks for implementing probabilistic teleportation of a
two-particle state, a three-particle state and a four-particle state
 of general form, built from single qubit gates, two-qubit controlled-not gates, Von Neumann
measurement and classically controlled operations.
\end{abstract}


\maketitle

\section{Introduction  }

Quantum teleportation, originally proposed by Bennett et al,  is one of the most striking progress of quantum
information theory [1].  It allows the transmission of an unknown qubit state from a sender "Alice" to a
spatially distant receiver "Bob" via a quantum channel with the aid of some classical communication. It will be
useful in quantum computers [2, 3].  It can be used to transmit information reliably in noisy situations where a
message would otherwise be degraded, and to transfer information from fleeting or hard-to-control carriers to
particles more suitably for permanent storage. Moreover, it has application in quantum cryptography and quantum
dense coding [4-8].

 At present, teleportation has been generalized to
 many cases [9-15] and demonstrated with the polarization photon [16]
and a single coherent mode of field [17] in the experiments [18, 19].   Due to the influence of environment, the
quantum channel composed of a pure entangled state is always taken
 to be nonmaximal. Therefore, it is of great interest to study the teleportatoin of an unknown
quantum state using a partly entangled state  as the quantum channel.  Shi et al and Lu et al proposed a protocol
for probabilistic teleportation of two-particle entangled state via a three-particle nonmaximally entangled state
[12, 14]. However, the form of two-particle entangled state is not a general form of two-particle state. In order
to conquer this limitation, Yan et al have generalized Shi's method to the two-particle  state of general form by
using partly  entangled four-particle pure state as the quantum channel [15].  In this scheme, an unknown
two-particle state of general form, whether entangled or not, can be transmitted from a sender to a receiver
through a partly  entangled four-particle pure state with certain probability. Unfortunately in their protocol
sixteen unitary transformations must be implemented. Obviously, it is not favorable to  the
 experimental realization of teleportation.

 Recently, Barenco et al showed that all unitary operations on arbitrarily
 many bits can be decomposed into the combinations of a set of one-bit quantum gates and two-bit Controlled-Not (CNOT)
 gates [20].  In terms of only single-qubit gates,
two-qubit CNOT gates, Von Neumann measurement and classically controlled operations, Liu et al presented quantum
logic networks for probabilistic teleportation of a single qubit and a two-particle entangled state, using a
partially entangled pair and a three-particle nonmaximally entangled state, respectively [21].  Clearly the
quantum logic networks for probabilistic teleportation will be important in realizing the teleportation scheme
in the experiment. Gao et al simplified the scheme of Ref.[15] first, then give a quantum logic network for
probabilistic teleportation of two-particle state of general form [22].

Since there are a great quantity of many particle quantum system in the real world, for example,  the quantum
computer is one, so it is important to study the teleportation of many particle state and deal with
teleportation problem in a comprehensive way. In this paper we discuss probabilistic teleportation of many
particle state.

This paper is organized as follows. We begin by introducing some notations and symbols in section 2.  In section
3  we develop an overall framework for probabilistically teleporting  an N-particle state of general form  via a
2N-particle nonmaximally entangled channel in a comprehensive way. The essential idea behind this framework may
be simplified in section 4 by considering the special cases when N=2, 3, 4. First, by means of the primitive
operations consisting of single-qubit gates, two-qubit CNOT gates , Von Neumann measurement and classically
controlled operations, we present an even simpler and more efficient quantum logic network than that of Ref.[22]
for probabilistically teleporting an unknown two-particle state of general form. Second, the  quantum networks
for probabilistic teleportation of a three-particle state  and a four-particle state of general form are also
exhibited.

\section{ Notation }

Some notations and symbols used throughout the paper are displayed here.
$$I=\left (
\begin{array} {cc}
1&0\\
0&1\\
\end{array}\right ); ~~~
\sigma_x=X=\left (
\begin{array} {cc}
0&1\\
1&0\\
\end{array}\right ); ~~~
\sigma_z=Z=\left (
\begin{array} {cc}
1&0\\
0&-1\\
\end{array}\right ); ~~~
H=\frac {1}{\sqrt 2}\left (\begin{array}{cc}
 1&1\\
 1&-1\\
 \end{array}\right).$$
 For any unitary operator
 \begin{equation}
 u=\left (
\begin{array} {cc}
u_{00}&u_{01}\\
u_{10}&u_{11}\\
\end{array}\right )
\end{equation}
 and $N\in \{0, 1, 2,...\}$, define the ($N$+1)-bit ($2^{N+1}$-dimensional) controlled operation $\Lambda_N(u)$
as
 \begin{equation}
 \Lambda_N(u)|x_1x_2\cdots x_N\rangle|y\rangle=|x_1x_2\cdots x_N\rangle u^{x_1x_2\cdots x_N}|y\rangle
 \end{equation}
for all  $x_1, x_2, \cdots, x_N, y\in \{0, 1\}$. Here $x_1x_2\cdots x_N$ in the exponent of $u$ means the
product of the bits $x_1, x_2, \cdots, x_N$. That is, the operator $u$ is applied to the last qubit if the first
$N$ qubits are all equal to one, and otherwise, nothing is done. Note that $\Lambda_0(u)$ is equated with $u$.
The $2^{(N+1)}\times 2^{(N+1)}$ matrix corresponding to $\Lambda _N(u)$ is
\begin{equation}
 \left (
\begin{array} {cccccccc}
1&&&&&\\
&1&&&&\\
&&\ddots&&&\\
&&&1&&\\
&&&&u_{00}&u_{01}\\
&&&&u_{10}&u_{11}
\end{array}\right )
\end{equation}
(where the basis states are lexicographically ordered, i.e., $|000\rangle, |001\rangle, \cdots, |111\rangle$).
$\Lambda_N(u)$ maps $|x_1x_2\cdots x_N\rangle|y\rangle$ to $|x_1x_2\cdots x_N\rangle|(\wedge_{k=1}^Nx_k)\oplus
y\rangle$ ( $\wedge_{k=1}^Nx_k$ denotes the AND of the Boolean variables {$x_k$}); that is, if the first $N$
qubits are all set to $|1\rangle$ then the last qubit  is flipped, otherwise the last qubit is left alone.

\section{Teleportation of a general  $N$-particle state }

In this section, we propose  the scheme for probabilistic teleportation of an $N$-particle state of general
form.

 Suppose that
Alice is to deliver an unknown normalized $N$-particle state of general form
\begin{equation}
\begin{array}{l}
|\phi\rangle_{123\cdots N}=x_0|000\cdots 00\rangle_{123\cdots N}+x_1|000\cdots 01\rangle_{123\cdots
N}+x_2|000\cdots 10\rangle_{123\cdots N}\\~~~~~~~~~~~~~~+ \cdots +x_{2^N-1}|111\cdots 11\rangle_{123\cdots
N}\\~~~~~~~~~~~ =\sum\limits_{i\in \{0, 1\}^N}x_i|i\rangle
\end{array}
\end{equation}
to a remote receiver Bob via a normalized quantum channel of a nonmaximally entangled state of $2N$ particles
\begin{equation}
\begin{array}{l}
|\phi\rangle_{N+1, N+2, \cdots, 3N}=\sum\limits_{i\in \{0, 1\}^N}y_i|ii\rangle.
\end{array}
\end{equation}
Here $x_i$'s are arbitrary complex numbers satisfying $\sum_{i=0}^{2^N-1}|x_i|^2=1$,  the notion '$\{0, 1\}^N$'
means 'the set of strings of length $N$ with each letter being either zero or one', and $y_0$ is the smallest of
real numbers $y_0$, $y_1$, $y_2$, $\cdots$, $y_{2^N-1}$. The combined state $|\Psi\rangle_{1, 2, 3, \cdots,
3N}=|\phi\rangle_{123\cdots N}|\phi\rangle_{N+1, N+2, \cdots, 3N}$ is the state of the total system. We use the
usual convention that the first $2N$ particles 1, 2, $\cdots$, $2N$ ( on the left ) belong to Alice, the other
$N$ particles $2N+1$, $2N+2$, $\cdots$, $3N$ belong to Bob. The sender Alice and the receiver Bob together
generated and shared the nonmaximally entangled state of Eq.(5), that is, Alice's last $N$ qubits and Bob's
qubits start out in a state of the form Eq.(5). Alice sends her qubits 1 and $N+1$, 2 and $N+2$, $\cdots$, $N$
and $2N$ through a CNOT gate respectively, and subsequently sends each of her first $N$ qubits through a
Hadamard gate. Then she measures the $2N$ qubits on her possession and transmits this information to Bob over a
classical communication channel. The post-measurement state of Bob's qubits will end up in $4^N$ possible
unnormalized states $|\psi_k\rangle_{2N+1, 2N+2, \cdots, 3N}$ ( $k=0, 1, 2, \cdots, 4^N-1$ ), one of which is
$|\psi_0\rangle_{2N+1, 2N+2, \cdots, 3N}=\sum\limits_{i\in \{0, 1\}^N}x_iy_i|i\rangle$.
 Depending on Alice's classical message, Bob knows exactly in which one of the $4^N$ states his qubits is. In
 order to achieve teleportation, Bob  needs to recover the original state $|\phi\rangle_{1,2,\cdots,N}$ at his
 side from the unnormalized states of his N particles.
 He introduces an auxiliary particle $a$ with the initial state $|0\rangle_a$
  and performs a collective
 unitary transformation
\begin{equation}
 U_N=\left (
\begin{array} {ccccc}
I&&&&\\
&u_1&&&\\
&&u_2&&\\
&&&\ddots&\\
&&&&u_{2^N-1}\\
\end{array}\right )
\end{equation}
on the state of particles $2N+1$, $2N+2$, $\cdots$, $3N$  and $a$, where
$$u_i=\left (
\begin{array} {cc}
\frac {y_0}{y_i}&-\sqrt {1-\frac {y_0^2}{y_i^2}}\\
\sqrt {1-\frac {y_0^2}{y_i^2}}&\frac {y_0}{y_i}\\
\end{array}\right ); ~~ i=1, 2, \cdots, 2^N-1.$$
If the measurement outcome on the auxiliary particle made by Bob is $|1\rangle_a$ , the teleportation fails.
While the measurement result is $|0\rangle_a$, the state of  Bob's particles $2N+1$, $2N+2$, $\cdots$, $3N$ is
in one of $4^N$ states $|\varphi_k\rangle_{2N+1, 2N+2, \cdots, 3N}$ not containing $y_k$ ( $k=0, 1, 2, \cdots,
4^N-1$ ), and Bob can 'fix up' his state, recovering  $|\phi\rangle_{123\cdots N}$, by the appropriate quantum
gates.

The unitary operator $U_N$ may be written  as{\small \begin{equation}
\begin{array}{l}
 U_N=(X^{\otimes(N-1)}\bigotimes I\bigotimes
I)\Lambda_N(u_1)(I^{\otimes(N-2)}\bigotimes X^{\otimes 2}\bigotimes I)\Lambda_N(u_2)(I^{\otimes(N-1)}\bigotimes
X\bigotimes I)\Lambda_N(u_3)(I^{\otimes(N-3)}\bigotimes
\\~~~~~~X^{\otimes3}\bigotimes I)\Lambda_N(u_4)(I^{\otimes(N-1)}\bigotimes X\bigotimes I)\Lambda_N(u_5)
(I^{\otimes(N-2)}\bigotimes X^{\otimes 2}\bigotimes I)\Lambda_N(u_6)(I^{\otimes(N-1)}\bigotimes X\bigotimes
I)\Lambda_N(u_7)\\~~~~~~(I^{\otimes(N-4)}\bigotimes X^{\otimes 4}\bigotimes
I)\Lambda_N(u_8)(I^{\otimes(N-1)}\bigotimes X\bigotimes I)\Lambda_N(u_9)(I^{\otimes(N-2)}\bigotimes X^{\otimes
2}\bigotimes I)\Lambda_N(u_{10})(I^{\otimes(N-1)}\bigotimes
\\~~~~~~X\bigotimes I)\Lambda_N(u_{11})(I^{\otimes(N-3)}\bigotimes
X^{\otimes3}\bigotimes I)\Lambda_N(u_{12})(I^{\otimes(N-1)}\bigotimes X\bigotimes I)\Lambda_N(u_{13})
(I^{\otimes(N-2)}\bigotimes X^{\otimes 2}\bigotimes I)\\~~~~~~\Lambda_N(u_{14})(I^{\otimes(N-1)}\bigotimes
X\bigotimes I) \Lambda_N(u_{15})(I^{\otimes(N-5)}\bigotimes X^{\otimes 5}\bigotimes
I)\Lambda_N(u_{16})(I^{\otimes(N-1)}\bigotimes X\bigotimes
I)\Lambda_N(u_{17})\\~~~~~~(I^{\otimes(N-2)}\bigotimes X^{\otimes 2}\bigotimes
I)\Lambda_N(u_{18})(I^{\otimes(N-1)}\bigotimes X\bigotimes I)\Lambda_N(u_{19})(I^{\otimes(N-3)}\bigotimes
X^{\otimes3}\bigotimes I)\Lambda_N(u_{20})(I^{\otimes(N-1)}\\~~~~~~\bigotimes X\bigotimes I)\Lambda_N(u_{21})
(I^{\otimes(N-2)}\bigotimes X^{\otimes 2}\bigotimes I)\Lambda_N(u_{22})(I^{\otimes(N-1)}\bigotimes X\bigotimes
I)\Lambda_N(u_{23})(I^{\otimes(N-4)}\bigotimes X^{\otimes 4}\bigotimes
I)\\~~~~~~\Lambda_N(u_{24})(I^{\otimes(N-1)}\bigotimes X\bigotimes
I)\Lambda_N(u_{25})(I^{\otimes(N-2)}\bigotimes X^{\otimes 2}\bigotimes
I)\Lambda_N(u_{26})(I^{\otimes(N-1)}\bigotimes X\bigotimes
I)\Lambda_N(u_{27})\\~~~~~~(I^{\otimes(N-3)}\bigotimes X^{\otimes3}\bigotimes
I)\Lambda_N(u_{28})(I^{\otimes(N-1)}\bigotimes X\bigotimes I)\Lambda_N(u_{29}) (I^{\otimes(N-2)}\bigotimes
X^{\otimes 2}\bigotimes I)\Lambda_N(u_{30})(I^{\otimes(N-1)}\\~~~~~~\bigotimes X\bigotimes I)
\Lambda_N(u_{31})(I^{\otimes(N-6)}\bigotimes X^{\otimes 6}\bigotimes
I)\Lambda_N(u_{32})(I^{\otimes(N-1)}\bigotimes X\bigotimes I)\Lambda_N(u_{33})(I^{\otimes(N-2)}\bigotimes
X^{\otimes 2}\bigotimes I)\\~~~~~~\cdots \Lambda_N(u_{2^N-3})(I^{\otimes(N-2)}\bigotimes X^{\otimes 2}\bigotimes
I)\Lambda_N(u_{2^N-2})(I^{\otimes(N-1)}\bigotimes X\bigotimes I)\Lambda_N(u_{2^N-1}).
\end{array}
\end{equation}}
Here $I^{\otimes (N-k)}$ and $X^{\otimes k}$ denote the parallel action of $N-k$ unity operators and $k$ NOT
gates, respectively.
 For saving space, we do not give an implementation of $U_N$ in terms of one and two qubit operations and
 also do not depict out the quantum circuit illustrating the
teleportation procedure of
 the $N$-particle state of general form via a shared $2N$-particle nonmaximally entanglement. But in what follows
  we  give the special cases when N=2, 3, 4.

\section{Teleportation of a general two-particle, three-particle and four-particle state}

In this section we  construct the quantum logic circuit implementing the probabilistic  teleportation process of
a two-particle, three-particle and four-particle state of general form. That will make the scheme of
probabilistic teleportation more concrete.

In Ref.[22] we discussed the case N=2. Here we take
\begin{equation}
U_2=\left(
\begin{array}{cccc}
I&&&\\
&u_1&&\\
&&u_2&\\
&&&u_3\\
\end{array}\right),
\end{equation}
which can also realize the teleportation with the same probability of success as in Ref.[22].
 $U_2$ can be expressed as
\begin{equation}
\begin{array}{l}U_2=\Lambda_2(u_3)(I\bigotimes X \bigotimes I)\Lambda_2(u_2)(X\bigotimes X\bigotimes
I)\Lambda_2(u_1)(X\bigotimes I\bigotimes I).
 \end{array}
\end{equation}
The gate array of $U_2$ is depicted in Fig.1.  By the technique for simulating $\Lambda_2(u)$ gates given by
Barenco et al and the Eq.(9) we derive
\begin{equation}
\begin{array}{l}U_2=(I\bigotimes I\bigotimes
A_3)(C^{5a}\bigotimes I^6) (I\bigotimes I\bigotimes B_3)(C^{5a}\bigotimes I^6)(C^{56}\bigotimes I)(I\bigotimes
I\bigotimes B_3)(I\bigotimes C^{6a})(I\bigotimes I\bigotimes A_3)\\~~~~~~ (I\bigotimes C^{6a})(C^{56}\bigotimes
I)(I\bigotimes I\bigotimes A_3)(I\bigotimes C^{6a})(I\bigotimes I\bigotimes B_3)(I\bigotimes C^{6a})(I\bigotimes
X \bigotimes I)(I\bigotimes I\bigotimes A_2)\\~~~~~~ (C^{5a}\bigotimes I^6) (I\bigotimes I\bigotimes
B_2)(C^{5a}\bigotimes I^6)(C^{56}\bigotimes I)(I\bigotimes I\bigotimes B_2)(I\bigotimes
 C^{6a})(I\bigotimes I\bigotimes A_2)(I\bigotimes C^{6a})\\~~~~~~ (C^{56}\bigotimes I)(I\bigotimes I\bigotimes A_2)(I\bigotimes C^{6a})
(I\bigotimes I\bigotimes B_2)(I\bigotimes C^{6a})(X\bigotimes X\bigotimes I)(I\bigotimes I\bigotimes
A_1)(C^{5a}\bigotimes I^6)\\~~~~~~ (I\bigotimes I\bigotimes B_1)(C^{5a}\bigotimes I^6)(C^{56}\bigotimes I)
(I\bigotimes I\bigotimes B_1) (I\bigotimes C^{6a})(I\bigotimes I\bigotimes A_1)(I\bigotimes
C^{6a})(C^{56}\bigotimes I)\\~~~~~~(I\bigotimes I\bigotimes A_1)(I\bigotimes C^{6a})(I\bigotimes I\bigotimes
B_1)(I\bigotimes C^{6a})(X\bigotimes I\bigotimes I)
\end{array}
\end{equation}
which is the decomposition constructed out of only two-qubit CNOT gate $\Lambda_1(X)$ along with a set of
one-qubit operations ( of the form $\Lambda_0(u)$ ). Here we assume $\cos\frac {\theta_1}{2}=\frac {y_0}{y_i}$,
  then $u_i=R_y(-\theta_i)=\left (
\begin{array} {cc}
\cos\frac {\theta_i}{2}&-\sin\frac {\theta_i}{2}\\
\sin\frac {\theta_i}{2}&\cos\frac {\theta_i}{2}\\
\end{array}\right )$, $A_i=R_y(-\frac{\theta_i}{4})$ and $B_i=R_y(\frac{\theta_i}{4})$ for $i=1,2,3$. In Eq.(10) $C^{ij}$
are the two-qubit CNOT gates $\Lambda_1(X)$ of control qubit $i$ and target qubit $j$ ( $i, j=5, 6, a$ ) and
$I^6$ means the unity operation on particle 6.

In this way, only a collective unitary operation $U_2$  is required to achieve the probabilistic teleportation
of an unknown  general two-particle quantum state $|\phi\rangle_{12}$, which, with regard to quantum gate array,
is simpler and more efficient than the scheme in Ref.[22] and even greater simplification than the scheme in
Ref.[15] where sixteen different collective unitary operations are needed for the same task.

For the case N=3, under the computational basis $|0000\rangle_{789a}, |0001\rangle_{789a}, |0010\rangle_{789a},
\cdots, |1111\rangle_{789a}$, Bob performs a unitary operation
\begin{equation}
 U_3=\left (
\begin{array} {ccccc}
I&&&&\\
&u_1&&&\\
&&u_2&&\\
&&&\ddots&\\
&&&&u_7\\
\end{array}\right)
\end{equation}
on the state of particles 7, 8, 9 and $a$.

If the measurement outcome made by Bob on the auxiliary qubit $a$ is $|1\rangle_a$, the teleportation fails.
However, if the measurement result is $|0\rangle_a$, the state of Bob's three  particles 7, 8 and 9 is one of
sixty-four states $|\varphi_0\rangle_{789}$,  $|\varphi_1\rangle_{789}$, $|\varphi_2\rangle_{789}$, $\cdots$,
$|\varphi_{63}\rangle_{789}$. There are no $y_i$ in the coefficients of these states, for instance, one of these
sixty-four states is
\begin{equation}
\begin{array}{l}
|\varphi_3\rangle_{789}=x_0|100\rangle_{789}+x_1|101\rangle_{789}+x_2|110\rangle_{789}+x_3|111\rangle_{789}-
x_4|000\rangle_{789}-x_5|001\rangle_{789}\\
~~~~~~~~~~~~~-x_6|010\rangle_{789}-x_7|011\rangle_{789}.
\end{array}
\end{equation}
The teleportation can be successfully achieved with the classical information from Alice and  corresponding
unitary operations ( $I$, $\sigma_x$, $\sigma_y$, $\sigma_z$ ) which are easily designed on the particles 7, 8
and 9. For example Bob can 'fix up' $|\varphi_3\rangle_{789}$ into the initial state of particles 1, 2 and 3 by
applying first a $\sigma_x$ and then a $\sigma_z$ gate on the particle 7.

The unitary operator $U_3$  can be written
\begin{equation}
\begin{array}{l}
U_3=(X\bigotimes X\bigotimes I\bigotimes I)\Lambda_3(u_1)(I\bigotimes X\bigotimes X\bigotimes
I)\Lambda_3(u_2)(I\bigotimes I\bigotimes X\bigotimes I)\Lambda_3(u_3)(X\bigotimes X\bigotimes X\bigotimes
I)\\
~~~~~~\Lambda_3(u_4)(I\bigotimes I\bigotimes X\bigotimes I)\Lambda_3(u_5)(I\bigotimes X\bigotimes X\bigotimes
I)\Lambda_3(u_6)(I\bigotimes I\bigotimes X\bigotimes I)\Lambda_3(u_7).
\end{array}
\end{equation}
With the help of the work done by Barenco et al [20], it is not difficult for us to give the decomposition of
$U_3$ using basic operations ( $\Lambda_0(u)$ and $\Lambda_1(X)$ ). For saving space, we will not exhibit it
here.  The sequence of the gate $U_3$ is illustrated in Fig.2.

The quantum circuit for the teleportation of  an unknown general three-particle quantum state
$|\phi\rangle_{123}$ is presented in Fig.3.

Here $T_1$, $T_2$, $\cdots$ ,$T_7$ are single-qubit rotation transformations. We can carefully choose $T_1$,
$T_2$, $\cdots$ ,$T_7$ to make the quantum state of the particles 4, 5, 6, 7, 8 and 9 to be the quantum channel
$|\phi\rangle_{456789}$ by the first dash line. If the measurement result on the auxiliary
 qubit at the output state is $|0\rangle_a$, the teleportation is
 successful with the final state of the particle 7, 8 and 9 being
 reconstructed as the initial state to be teleported.

Note that a controlled unitary operation acting on any number of qubits followed by the measurement of the
control qubit can be replaced by the measurement of the control qubit preceding  the controlled operation [21].
Therefore, Fig.3 can be re-expressed as Fig.4.

In Fig.4, the controlled operation can be realized locally by Bob depending on the results of the six
measurements performed by Alice  on her own qubits. If and only if the outcome of Alice's measurement is 1, Bob
can execute the Controlled-Not $X$ or controlled-$Z$. If the measurement result of the auxiliary particle is
$|1\rangle_a$, the teleportation fails. If the result is $|0\rangle_a$, the final state of particles 7, 8 and 9
at Bob's side will be collapsed into $|\phi\rangle_{789}=\sum_{i=0}^7x_i|i\rangle_{789}$, which is the desired
state. That is to say, if the measurement outcome on the state of the auxiliary qubit is $|0\rangle_a$, perfect
teleportation is accomplished.

For the case N=4, the unitary operator
\begin{equation}
\begin{array}{l}
U_4=(X\bigotimes X\bigotimes X\bigotimes I\bigotimes I)\Lambda_4(u_1)(I\bigotimes I\bigotimes X\bigotimes
X\bigotimes I)\Lambda_4(u_2)(I\bigotimes I\bigotimes I\bigotimes X\bigotimes I)\Lambda_4(u_3)\\~~~~~~(I\bigotimes
X\bigotimes X\bigotimes X\bigotimes I)\Lambda_4(u_4)(I\bigotimes I\bigotimes I\bigotimes X\bigotimes
I)\Lambda_4(u_5)(I\bigotimes I\bigotimes X\bigotimes X\bigotimes I)\Lambda_4(u_6)\\
~~~~~~(I\bigotimes I\bigotimes I\bigotimes X\bigotimes I)\Lambda_4(u_7) (X\bigotimes X\bigotimes X\bigotimes
X\bigotimes I)\Lambda_4(u_8)(I\bigotimes I\bigotimes I\bigotimes X\bigotimes I)\Lambda_4(u_9)\\
~~~~~~(I\bigotimes I\bigotimes X\bigotimes X\bigotimes I)\Lambda_4(u_{10})(I\bigotimes I\bigotimes I\bigotimes
X\bigotimes I)\Lambda_4(u_{11})(I\bigotimes X\bigotimes X\bigotimes X\bigotimes I)\Lambda_4(u_{12})\\
~~~~~~ (I\bigotimes I\bigotimes I\bigotimes X\bigotimes I)\Lambda_4(u_{13})(I\bigotimes I\bigotimes X\bigotimes
X\bigotimes I)\Lambda_4(u_{14})(I\bigotimes I\bigotimes I\bigotimes X\bigotimes I)\Lambda_4(u_{15}).
\end{array}
\end{equation}
 Fig.5 gives an implementation of $U_4$.  With the result provided by Barenco et al regarding how to
simulate a general controlled operation $\Lambda_4(u)$ from only single-qubit gates  $\Lambda_0(u)$ and
two-qubit CNOT gates $\Lambda_1(X)$, $U_4$ can be built from just this gate set. For saving space, we will not
exhibit it here.

 The teleportation of  an unknown general four-particle quantum state
$|\phi\rangle_{1234}$ may be implemented using  quantum circuit shown in Fig.6.

Here $T_1$, $T_2$, $\cdots$ ,$T_{15}$ are single-qubit rotation transformations. We can properly choose $T_1$,
$T_2$, $\cdots$ ,$T_{15}$ to make the quantum state of the particles 5, 6, 7, 8, 9, 10, 11 and 12 to be the
quantum channel $|\phi\rangle_{56789101112}$ by the first dash line. If the measurement result on the auxiliary
 qubit at the output state is $|0\rangle_a$, the teleportation is
 successful with the final state of the particle 9, 10, 11 and 12 being
 reconstructed as the initial state to be teleported.

Clearly, Fig.6 can be re-expressed as Fig.7.

In Fig.7 if the measurement result of the auxiliary particle is $|1\rangle_a$, the teleportation fails. If the
result is $|0\rangle_a$, the final state of particles 9, 10, 11 and 12 at Bob's side will be collapsed into
$|\phi\rangle_{9,10,11,12}=\sum_{i=0}^{15}x_i|i\rangle_{9,10,11,12}$, which is the desired state. That is to
say, if the measurement outcome on the state of the auxiliary qubit is $|0\rangle_a$, perfect teleportation is
accomplished.

In summary, a protocol of probabilistic teleporting an unknown general $N$-particle state by a shared
$2N$-particle nomaximally entangled state is presented. As special cases, the quantum circuits for probabilistic
teleporting a three-particle state and a four-particle state of general form  via a six-particle and an
eight-particle nonmaximally entangled state are constructed respectively. By means of an auxiliary qubit and a
unitary operation, a simpler scheme than that in Ref.[15] and a more efficient quantum circuit than that in
Ref.[22] for probabilistic teleportation of a two-particle state of general form through a shared four-particle
nonmaximally entangled state were exhibited. It deserves to be mentioned that as long as the sender and all the
receivers initially share the form of the entanglement, as specified in our paper, an unknown $N$-particle
general state can be probabilistically teleported to many
spatially receivers.\\

\begin{acknowledgements}
 This work was supported by National Natural Science Foundation of
China under Grant No. 10271081 and Hebei Natural Science Foundation under Grant No. 101094.
\end{acknowledgements}

\footnotesize
\begin{tabbing}
xxxxx\=\kill ~1. C.H. Bennett, G. Brassard, C. Crepeau, R. Jozsa,
A. Peres, and W.K. Wootters,  Phys. Rev. Lett.
  {\bf 70}, \\
  ~~~~~1895 (1993).\\
~2. J.I. Cirac and P. Zoller, Phys. Rev. Lett. {\bf 74}, 4091 (1995).\\
~3. A. Barenco, D. Deutsch, A.K. Ekert, and R. Jozsa,  Phys. Rev. Lett. {\bf 74}, 4083 (1995).\\
~4. A.K. Ekert,  Phys. Rev. Lett.  {\bf 67}, 661  (1991).\\
~5. C.H. Bennett,  Phys. Rev. Lett.  {\bf 68},  3121 (1992).\\
~6. Y. Zhang, L. Deng, M. Mao, and L.E. Ding, Chin. Phys. Lett. {\bf 15}, 238 (1998).\\
~7. B.S. Shi  and G.C. Guo,  Chin. Phys. Lett.  {\bf 14}, 521 (1997).\\
~8. C.H. Bennett  and  S.J. Wiesner,  Phys. Rev. Lett.  {\bf 69}, 2881 (1992).\\
~9. M. Ikram, S.Y. Zhu, and M.S.  Zubairy, Phys. Rev. A{\bf 62}, 022307 (2000). \\
10. W.L. Li, C.F. Li, and G.C. Guo, Phys. Rev. A{\bf 61},  034301 (2000).\\
11. V.N. Gorbachev  and A.I. Trubilko, J. Exp. Theor. Phys.  {\bf 91}, 894 (2000).\\
12. H. Lu and  G.C. Guo, Phys. Lett.   A{\bf 276}, 209 (2000).\\
13. B. Zeng, X.S. Liu, Y.S. Li, and G.L. Long, Commun. Theor. Phys. {\bf 38}, 537 (2002).\\
14. B.S. Shi, Y.K. Jiang, and G.C. Guo,  Phys. Lett. A{\bf 268}, 161 (2000).\\
15. F.L. Yan, H.G. Tan, and L.G. Yang, Commun. Theor. Phys. {\bf 37}, 649 (2002).\\
16. D. Bouwmeester, J.W. Pan, K. Mattle, M. Eible, H. Weinfurter, and A. Zeilinger,  Nature  {\bf 390}, 575 (1997).\\
17. A. Furusawa, J.L. Sorensen, S.L. Brawnstein, C.A. Fuchs, H.J. Kimble, and E. Polizk,  Science  {\bf 282}, 706 (1998).\\
18. D. Boschi,  S. Branca, F. DeMartini, L. Harely, and S. Popescu, Phys. Rev. Lett.   {\bf 80}, 1121 (1998).\\
19. M.A. Nielsen, E. Knill, and R. Laflamme,  Nature    {\bf 396}, 52 (1998).\\
20. A. Barenco, C.H. Bennett, R. Cleve, D.P. DiVincenzo, N.
Margolus, P. Shor, T. Sleator, J.A. Smolin, and
\\~~~~~H. Weinfurter,
 Phys. Rev. A{\bf 52}, 3457 (1995).\\
21. J.M. Liu, Y.S. Zhang, and G.C. Guo, Chinese Physics {\bf 12}, 251 (2003).\\
22. T. Gao, Z.X. Wang, and F.L. Yan, Chin. Phys. Lett. {\bf 20}, 2094 (2003).
\end{tabbing}

\end{document}